# Growth of Epitaxial MgB$_2$ Thick Films with Columnar Structures by Using HPCVD**

By *Won Kyung Seong, Ji Yeong Huh, Won Nam Kang,* * *Jeong-Woon Kim, Yong-Seung Kwon, Nam-Keun Yang, and Je-Geun Park*

Epitaxial MgB$_2$ thick films were grown on Al$_2$O$_3$ substrates at 600 °C by using the hybrid physical chemical vapor deposition (HPCVD) technique. In order to obtain a high magnesium vapor pressure around the substrates, we used a special susceptor having a susceptor cap and achieved a very high growth rate of 0.17 µm/min. Hexgonal-shaped columnar structures were observed by cross-sectional and planar view transmission electron microscope (TEM) images. For the 1.7-µm-thick film, the $T_c$ was observed to be 40.5 K with a $J_c$ of $1.5 \times 10^6$ A/cm$^2$ at 30 K. The vortex pinning mechanism by intercolumnar boundaries will be discussed.

Keywords: MgB$_2$ thick film, Columnar grains, Critical current, HPCVD, TEM

[*]   Prof. Won Nam Kang, Mr. Won Kyung Seong, Miss Ji Young Huh, Mr. Jeong-Woon Kim Kim, Prof. Yong-Seung Kwon, Mr. Nam-Keun Yang, Prof. Je-Geun  Park
       BK21 Physics Division and Department of Physics, Sungkyunkwan University
       Suwon 440-746 (Korea)
       E-mail: wnkang@skku.edu

[**] This work was supported by the Korea Research Foundation grant funded by the Korean Government (MOEHRD) (KRF-2005-005-J11902 & KRF-2006-312-C00130) and by the Korea Science and Engineering Foundation (KOSEF) grant funded by the Korea Government (MOST) (R01-2005-000-11001-0). JGP was supported by the Korea Research Foundation (Grant No. 2005-C00153) and the LG Yonam Foundation.

## 1. Introduction

The dependences of the critical current density ($J_c$) on temperature and magnetic field are the important aspect of superconducting materials for basic research and applications. Binary metallic $MgB_2$ with a high transition temperature of 39 K has attracted great interest in both fundamental studies and practical applications.[1,2] Some companies have commercially produced long $MgB_2$ wires from conventional powder by using a powder in tube method.[3] For high-temperature applications above 20 K, however, $MgB_2$ superconducting tapes or wires with high $T_c$ and $J_c$ at high temperature must be produced in long lengths. In addition, the phase purity and the high orientation of $MgB_2$ thick films are important factors that need some attention for producing superconducting tapes and wires. Many groups have tried to improve the quality of $MgB_2$. For example, $J_c$ values of $3 \times 10^5$ A/cm$^2$ at 20 K for the bulk,[4] $1.6 \times 10^7$ A/cm$^2$ at 15 K for a thin film,[5] $7.4 \times 10^6$ A/cm$^2$ at 5 K for a thick film,[6] $3 \times 10^5$ A/cm$^2$ at 4.2 K for a wire,[7] and $4 \times 10^5$ A/cm$^2$ at 4.2 K for a tape[8] have been demonstrated at zero field.

Several deposition procedures are used to fabricate superconducting $MgB_2$ films. As *ex-situ* techniques, two-step processes employing *ex-situ* annealing of deposited boron[2] or magnesium diboride[9] films at high temperatures have been used. Kang and co-workers optimized the *ex-situ* annealing temperature and time and fabricated thin films with a high $J_c$ of $\sim 10^7$ A/cm$^2$ at 15 K under zero field.[5] However, the *ex-situ* methods are usually very difficult, and a long process time is required to fabricate high-quality $MgB_2$ films. Zeng *et al*. developed an *in-situ* hybrid physical chemical vapor deposition method to grow high-quality $MgB_2$ film with a high $J_c$ of $\sim 10^7$ A/cm$^2$ at 5 K at zero field.[10] Thus, HPCVD should be a very useful technique for fabricating $MgB_2$ thick films.

In this paper, we report the fabrication of $MgB_2$ thick films by using HPCVD; the films showed epitaxial growth with columnar grain structures. The columnar grains were grown on a (0001) $Al_2O_3$ substrate perpendicular to the substrate surface, and the average grain sizes were observed to be $\sim 300$ and $\sim 400$ nm in diameter for 1.0-μm-thick and 1.7-μm-thick $MgB_2$ films, respectively. Different from high-$T_c$ superconductors,[3] not only are the grain boundaries in $MgB_2$ superconductors transparent to a current but they also contribute significantly to the strong pinning.[11] We found that the $J_c$ under a magnetic field was substantially enhanced by the columnar grain boundaries. At 30 K, the values of $J_c$ were $4 \times 10^6$ A/cm$^2$ and $1.5 \times 10^6$ A/cm$^2$ for 1.0-μm-thick and 1.7-μm-thick $MgB_2$ films, respectively.

## 2. Results and Discussion

Figure 1(a) shows the temperature dependence of the resistivity of the $MgB_2$ thick films. The residual resistivity ratio (RRR) of the thick films was $\sim 4$ for both the 1.0-μm-thick and the 1.7-μm-thick films. A magnified view in the temperature region from 38 to 42 K was plotted in Fig. 2(b), for the sake of clarity. The $T_c$'s of the $MgB_2$ thick films depended on the film thickness. The thicker film showed a higher $T_c$ than the thinner film (40.5 K vs 39.8 K). This is fairly consistent with earlier results [12], in which they reported a thickness dependence of $T_c$ in $MgB_2$ thin films grown on $Al_2O_3$ substrates by using HPCVD and observed the highest values of $T_c$, 40.5 K. Based on a X-ray diffraction (XRD) and Raman scattering experiments, they proposed that the enhancement of $T_c$ in $MgB_2$ films was due to a softening of a bond-stretching phonon mode with a biaxial tensile strain.

Figure 2(a) shows XRD pattern for the 1.7-μm-thick $MgB_2$ films. The lattice constant of the *c*-axis determined from the (000*l*) peaks is observed to be 3.513 Å, slightly smaller than the bulk value of 3.524 Å. According to the Bardeen-Cooper-Schrieffer (BCS) theory, $T_c$ depends on the electron-phonon interaction. The origin of the enhanced $T_c$ in our thick films probably comes from the increased *c*-axis electron-phonon interaction. Figure 2(b) shows the *φ*-scan parallel to the ($10\bar{1}1$) plane reflection of the $MgB_2$ films. The periodic peaks separated by 60° reveal the six-fold symmetry of the $MgB_2$ thick film, indicating the presence of epitaxial growth. The smaller periodic peaks indicate 30° rotational twinning between the $MgB_2$ thick film and the $Al_2O_3$ substrate.[10]

Figure 3(a) shows the zero-field-cooled (ZFC) and the field-cooled (FC) dc magnetization (*M*) curves for the $MgB_2$ thick films measured with 1 mT applied parallel to the *c*-axis. The ZFC curve shows a diamagnetic transition, as compared to the resistivity data. Figure 3(b) shows the magnetization (*M − H*) hysteresis loops for the 1.7-μm-thick and 1.0-μm-thick films measured by using a superconducting quantum interference devices (SQUIDs) at temperatures of 5 K to 30 K.

Figure 4(a) shows an a scanning electron microscope (SEM) image of the columnar grain structure on the surface for 1.7-μm- thick film. Figure 4(b) shows a planar-view, TEM, bright field image obtained at a ~1-μm depth from the film surface. Hexagonal grain structures with an average diameter of ~ 300 nm can be clearly seen. It should be noted that an average diameter of ~ 400 nm was observed from the SEM images for the 1.0-μm-thick film, which is not shown here. The grain boundaries between these columnar grains may work as effective pinning centers and lead to the high $J_c$ performance for field perpendicular to substrate surface.[13] Figure 4(c) shows a cross-sectional TEM bright-field image. The columnar grains were a direction perpendicular to the surface of the substrate. The in-plane selected area electron diffraction (SAED) pattern shows a zone axis of Z = [1000], as indexed in Fig. 4(d). These results indicate that the MgB$_2$ grains have been grown along the [0001] direction. Although most of the intercolumnar boundaries were parallel to the c-axes of the thick films, some twist-formed intercolumnar boundaries were observed on in-plane SAED pattern, as shown Fig. 4(e).

The $J_c$'s vs. magnetic field at 5 K and 30 K for MgB$_2$ thick films are shown in Fig. 5(a). The $J_c$ was calculated from the $M-H$ loops by Bean's critical state model ($J_c = 30\Delta M/r$), where $\Delta M$ is the height of $M-H$ loops and $r$ is the radius corresponding to the total area of the film surface.[5] At 30 K in zero field, the thinner (1.0 μm) film shows a much higher $J_c$ than the thicker (1.7 μm) film whereas the $J_c-H$ curves coincide at higher fields, suggesting that the grain connectivity of the thinner film is stronger than that of the thicker film. On the other hand, at 5 K, the two samples had similar values of $J_c$ ~$10^7$ A/cm$^2$ at low field, but the difference between the 1.7-μm-thick and the 1.0-μm-thick films increased dramatically with increasing applied magnetic field. The value of $J_c$ in the thicker film was one order higher in magnitude than that in the thinner film even at H = 4 T, implying that the thicker film had a much higher density of strong pinning centers with weak intercolumnar regions compared to the thinner films. These results suggest that the thicker films grown by using the HPCVD technique may be useful in fabricating MgB$_2$ tapes having high values of $J_c$ at high temperatures. For high-field applications, however, thicker films having smaller columnar grains should be fabricated in order to enhance the pinning-site density of the intercolumnar grain boundaries.

Figure 5(b) show the field dependence of the pinning force density, $F_p = J \times B$, for the 1.0-μm-thick and the 1.7-μm-thick films at 5 and 30 K, where $F_p$ is normalized by the maximum pinning force ($F_{p,max}$). The values of $F_p$ of the two samples measured at 30 K fell on one curve, which is consistent with the previous discussion on the field dependence of $J_c$. For the data measured at 5 K, the maximum peaks were observed at 0.7 T and 1.4 T for the 1.0-μm-thick and the 1.7-μm-thick films, respectively, indicating that the thicker film contained a higher pinning density than thinner one.

## 3. Conclusion

In summary, we have fabricated high-quality epitaxial MgB$_2$ thick films with c-axis oriented columnar structure by using HPCVD. The as-grown films on Al$_2$O$_3$ substrate showed a high $T_c$ of 40.5 K for the 1.7-μm-thick film. By using a special susceptor having a susceptor cap, we have achieved a very high growth rate of 0.17 μm/min, which is much higher value than that of the normal HPCVD process. We suggest that grain boundaries between columnar grains lead to the high $J_c$. Thus, the HPCVD process should be a promising candidate for practical applications, such as fabricating MgB$_2$ wires and tapes.

## 4. Experimental

*HPCVD system*: HPCVD system consists of a vertical quartz tube reactor having a bell shape with 65mm in inner diameter and 200 mm in height, inductively coupled heater and a load-rock chamber. Schematic diagram of this system was previously described in more details.[14]

*Susceptor*: Different from a typical susceptor used by Zeng et al.,[10] we employed a special susceptor with a susceptor cap in order to obtain a high magnesium vapor pressure around the substrate at low temperature, as shown in Fig. 6. With this special reactor, we were able to achieve very high growth rate of 0.17 μm/min at a low temperature below 600°C. Also, we used a stainless-steel susceptor, which is cheap and easy to manufacture and handle. The size of substrate holder is as large as 25.4 mm in diameter so that we are able to grow large area MgB$_2$ thin films. Temperature calibration was carried out by using type-K thermocouple. The calibrated temperatures of substrate and Mg crucible were 20 °C lower than the temperature reading by the control thermocouple placed inside the susceptor.

*Precursors and substrates*: Small Mg chips (99.999%) with 2 − 3 mm in diameter and B$_2$H$_6$ (5% in H$_2$) gas were used as Mg and B sources.

The substrates used in this study were (0001) $Al_2O_3$ with $0.5 \times 10 \times 10$ mm$^3$.

*HPCVD process*: After installing the susceptor with substrate and 2 g of Mg chips in the reactor, we purged the reactor for several times by flowing high purity Ar gas. The substrates were inductively heated up to 580 ~ 600 $^o$C under a reactor pressure of 200 Torr in ambient $H_2$; then, $H_2$ carrier gas and $B_2H_6$ (5% in $H_2$) reactive gas were introduced. The total flow rate of the gas mixture was 150 sccm, corresponding to a 1% $B_2H_6$ concentration in the $H_2$ gas. These process temperatures were chosen as optimal after few attempts to grow thin films at various substrate temperatures from 500 $^o$C to 760 $^o$C. With increasing the substrate temperatures, the film growth rate decreased because sticking coefficient of Mg decreased.

*Film characterization*: Film thickness was measured by using SEM (Nova 600, FEI) after cross-sectional cutting by focused ion-beam (FIB). The surface morphology of $MgB_2$ thin films were investigated by using SEM. The specimens for TEM (CM-30, Philips) were also prepared by FIB. Structural analyses were carried out by XRD (D8 discover, Bruker AXS) with Cu K$_\alpha$ X-ray source. The temperature dependences of magnetizations and the magnetization hysteresis (*M* − *H*)loops were measured by the magnetic property measurement system (XL-5S, Quantum design). For the electrical measurements, the films were cut into $2 \times 5$ mm$^2$ and the resistivity was measured by a conventional four-probe method.

**Figures and Figure Captions**

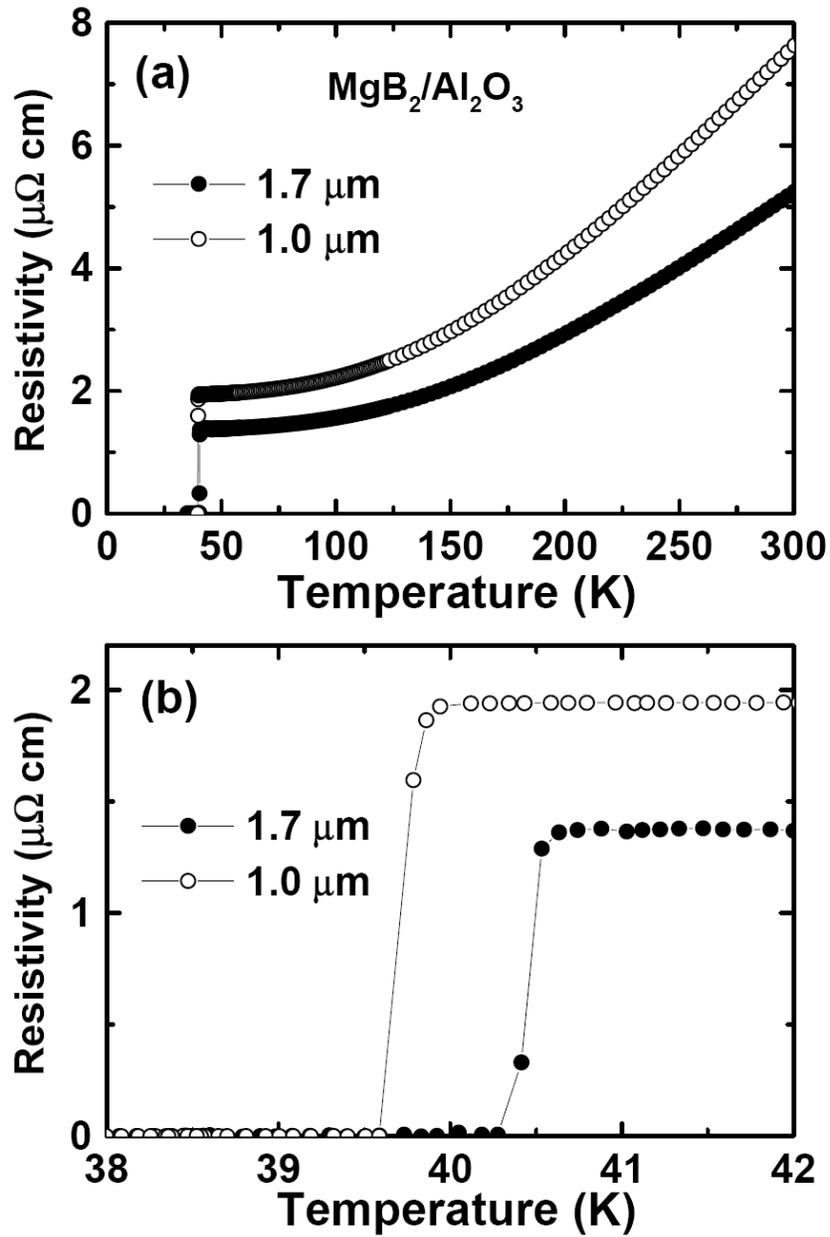

Fig. 1. (a) Resistivity versus temperature for 1.0-μm-thick and 1.7-μm-thick MgB$_2$ films on Al$_2$O$_3$ substrates, and (b) a magnified view in the temperature region from 38 to 42 K.

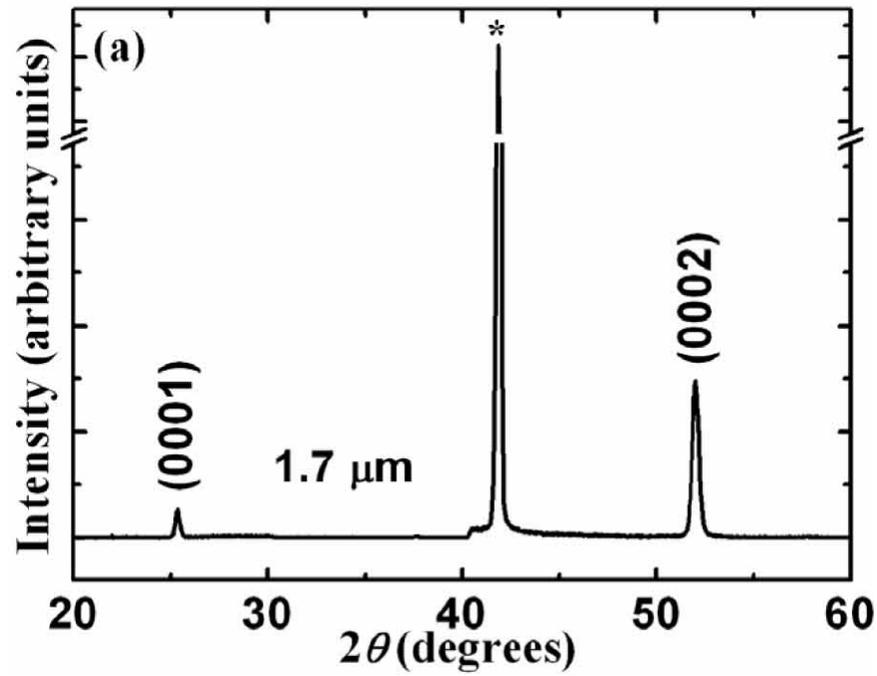

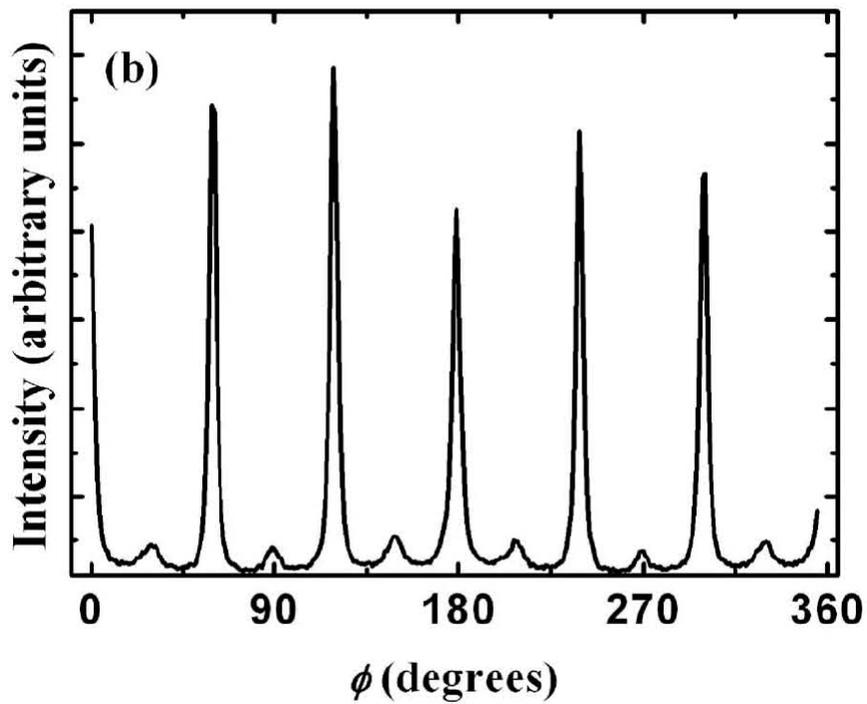

Fig. 2. XRD pattern for 1.7-μm-thick MgB$_2$ film: (a) $\theta$-$2\theta$ scan and (b) $\varphi$-scan of the (10$\bar{1}$1) MgB$_2$ reflection. The substrate peak is marked by an asterisk (*).

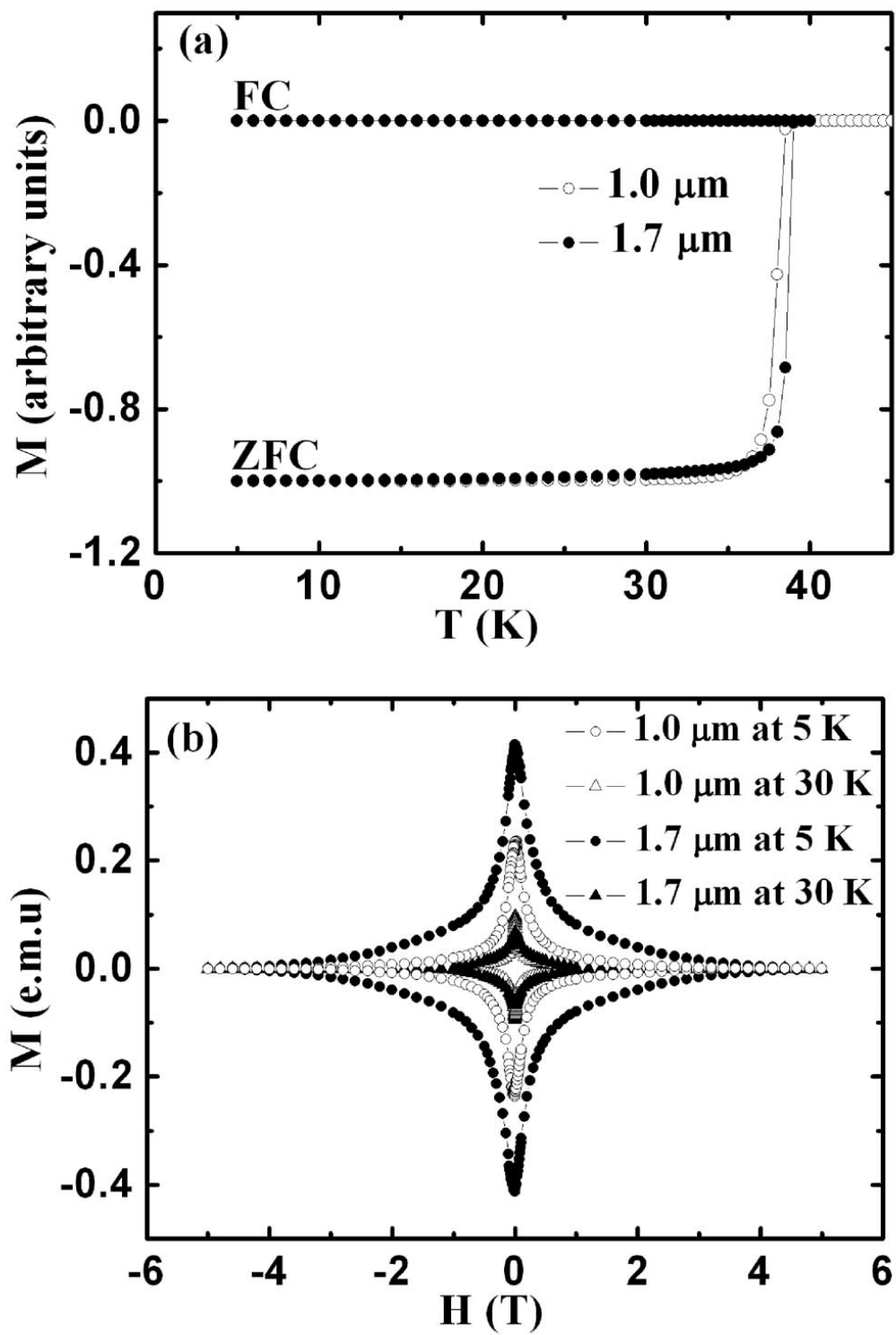

Fig. 3. (a) Temperature dependence of the magnetization at 1 mT for $MgB_2$ thick films (b) *M-H* loops at 5 K and 30 K for 1.0-μm-thick and 1.7-μm-thick $MgB_2$ films.

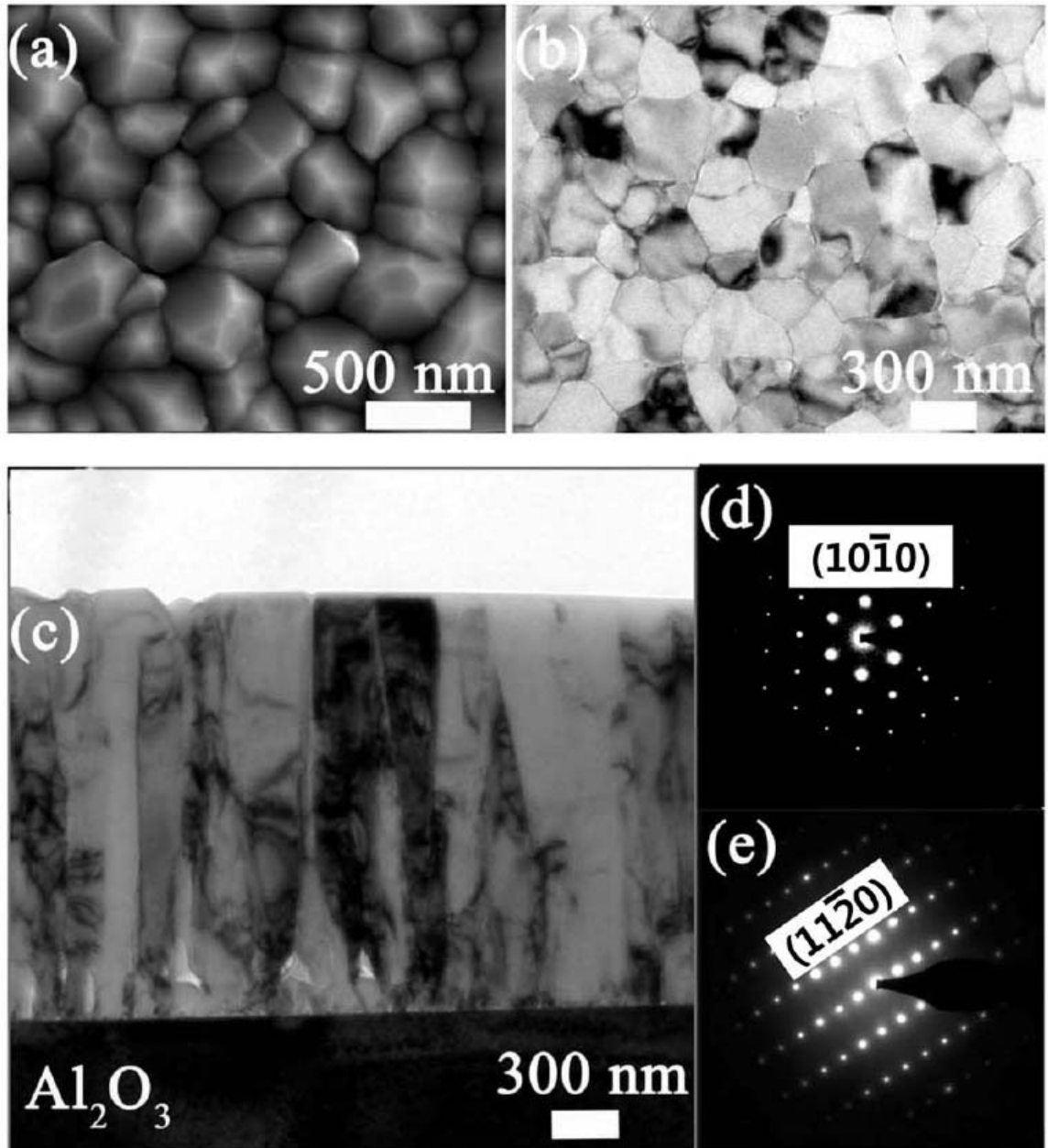

Fig. 4. (a) SEM image of the 1.7-μm-thick MgB$_2$ film surface. (b) In-plane view of TEM image for the 1.7-μm-thick MgB$_2$ film. (c) Cross-sectional TEM image for the 1.7-μm-thick MgB$_2$ film. (d) In-plane SAED image for the 1.7-μm-thick film. (e) Out-of-plane SAED image for the 1.7-μm-thick film.

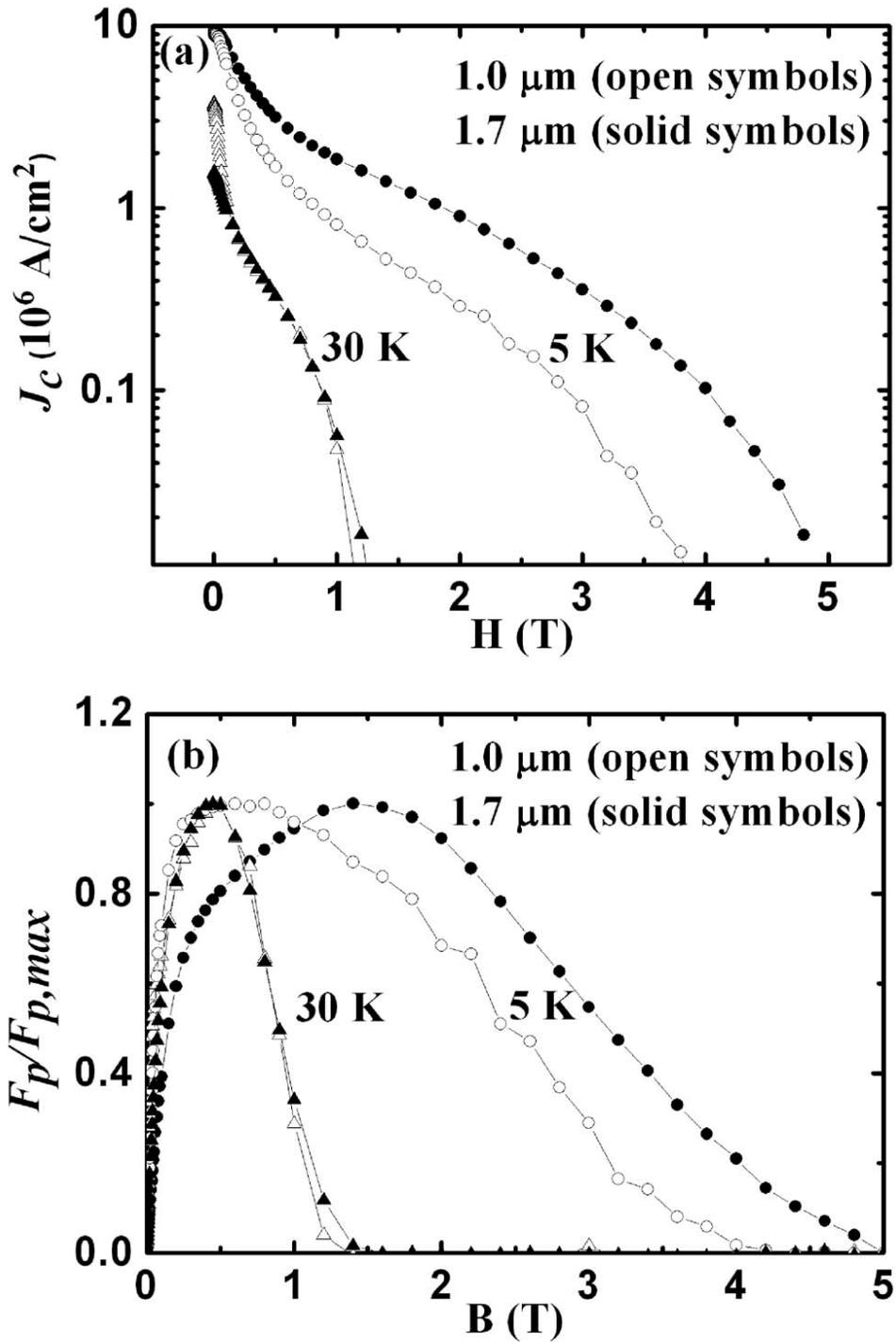

Fig. 5. (a) The field dependence of $J_c$ for 1.0-μm-thick and 1.7-μm-thick $MgB_2$ films at 5 K and 30 K. (b) Normalized pinning force ($F_p/F_{p,max}$) vs. magnetic field at 5 K and 30 K for 1.0 μm and 1.7 μm thick $MgB_2$ films.

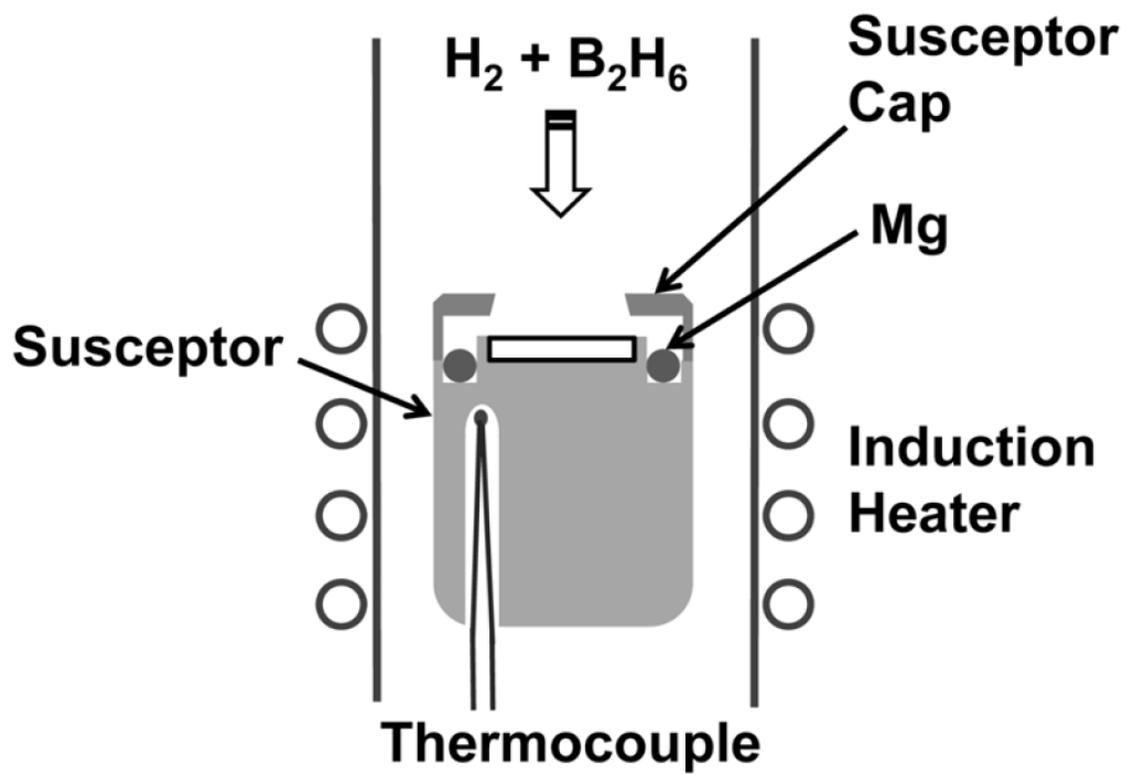

Fig. 6. Schematic diagram of HPCVD reactor and susceptor.